\documentclass[fleqn,twocolumn, numberedappendix]{openjournal} 

\usepackage{ulem}

\DeclareRobustCommand{\VAN}[3]{#2}
\let\VANthebibliography\thebibliography
\def\thebibliography{\DeclareRobustCommand{\VAN}[3]{##3}\VANthebibliography}

\usepackage{xcolor}

\usepackage{graphicx}	
\usepackage{amsmath}	
\usepackage{amssymb}	

\usepackage{hyperref}
\hypersetup{
    colorlinks = true,
    linkcolor = blue,
    citecolor = blue
    }
    
\usepackage[
        noabbrev,
        capitalise,
        nameinlink,
    ]
    {cleveref}
\usepackage{amsthm}

\begin{document}
\title{Observational implications of Cosmologically coupled black holes}


\author{Sohan Ghodla\textsuperscript{1}}\thanks{\textsuperscript{1} \texttt{sgho069@aucklanduni.ac.nz}}
\author{Richard Easther\textsuperscript{2}}\thanks{\textsuperscript{2} \texttt{r.easther@auckland.ac.nz}}
\author{M. M. Briel} 
\author{J.J. Eldridge} 
\affiliation{Department of Physics, 
University of Auckland 1010, 
Private Bag 92019, 
Auckland,
New Zealand}




\label{firstpage}

\begin{abstract}
 It was recently suggested that ``cosmologically coupled'' black holes with masses that increase in proportion to the volume of the Universe might constitute the physical basis of dark energy. We take this claim at face value and discuss its potential astrophysical  implications. We show that the gravitational wave emission in binary systems would be significantly enhanced so that the number of black hole mergers would exceed the observed rate by orders of magnitude, with typical masses much larger than those seen by the LIGO-Virgo-KAGRA network. Separately, if the mass growth happens at fixed angular momentum, the supermassive black holes in matter-deficient elliptical galaxies should be slowly rotating. Finally,  cosmological coupling would stabilize small black holes against Hawking radiation-induced evaporation.

\end{abstract}
\maketitle






\section{Introduction} \label{sec: intro}

It has been suggested that there is a tight coupling between the physics of black holes and the global properties of spacetime (e.g., \citealt{Faraoni:2007es,Guariento:2012ri,Maciel:2015dsh}), with the result that black hole masses may evolve with the expanding Universe. In particular, building on arguments presented by  \cite{Croker_Weiner:2019}, \cite{Farrah_2023a} recently claimed that this effect could account for the redshift dependence of the size of supermassive black holes (SMBH) at the centers of red-sequence elliptical galaxies.

A generic form for such a coupling is \citep{Croker_2021}
\begin{equation}
    m(a) = m_i \left(\frac{a}{a_i}\right)^k \quad a \geq a_i,
    \label{eq: Eqn from Farrah et al. for mass growth}
\end{equation}
where $a$ is the cosmic scale factor and $a_i$ is its value when the black hole forms with ``birth mass'' $m_i$. \cite{Farrah_2023} find that the redshift dependence of SMBH masses is best-fit by $k \approx 3$ and argue that the spatially averaged density of stellar-remnant black holes is thus (roughly) constant. Furthermore, they claim that the resulting negative pressure matches that of the dark energy in the concordance cosmological model and would account for the accelerated expansion of the present-day Universe. This claim has been critiqued both on theoretical grounds (e.g., \citealt{Parnovsky:2023, Wang:2023, Mistele:2023, cadoni2023cosmological}) and in relation to its observational implications \citep{Rodriguez:2023, Andrae:2023, amendola2023constraints}.

In this paper, we take the framework of \cite{Farrah_2023} at face value and consider its astrophysical implications. We assume that  stellar evolution is unchanged from standard expectations  and that  black holes form at the endpoints of the evolution of massive stars. We use \textsc{Bpass}, a binary star population synthesis tool \citep{Bpass2017, Bpass_2018}, to compute the binary black hole (BBH) formation rate as a function of redshift.  Given that black hole mergers occur on cosmological timescales, the mass growth between birth and eventual merger is typically substantial, which will alter the  merger dynamics of these binaries. We compute the source frame merger rates of BBHs after accounting for these effects, along with the expected detection rates in the LIGO-Virgo-KAGRA (LVK) detectors \citep{GWTC_3_population_stats}. 
The merger rate and the typical merger masses increase significantly relative to $k=0$ and would be in substantial conflict with observations.

We also consider the impact of a time-dependent mass on the spin of Kerr black holes. If the cosmological coupling does not add angular momentum, the spin parameter will decrease with time.   
Consequently, black holes situated in matter-deficient environments and (in particular) SMBH in red-sequence elliptical galaxies will generally exhibit lower spin values.

Lastly, we discuss the implications of the mass growth effect for hypothetical primordial black holes (PBHs) so that the mass loss through Hawking radiation competes with cosmologically driven mass growth. Given that PBHs presumably form at very high redshifts, any such objects are likely to be very massive in the present epoch.

The structure of this paper is as follows. In Section~\ref{sec: mergertime calculation}, we extend treatments of orbital decay via gravitational radiation to scenarios with time-varying masses. In Section~\ref{sec: merger_rates result}, we compute the expected population of merger events visible in terrestrial detectors, comparing the  $k=0$ and $k=3$ cases with observations. Section~\ref{sec: spin of cosmological coupled BH} examines the impact of mass variation on the angular momentum of black holes, and Section \ref{sec: minimum mass of primordial BHs} considers the survival of initially small PBH. We end with a brief discussion and summary of our results in Section~\ref{sec: Discussion}.

\section{Mergers with time-dependent masses} \label{sec: mergertime calculation}

The framework of \cite{Croker_Weiner:2019} is formulated within Einstein-Hilbert gravity -- there is no ``new physics'', nor any additional free parameters. Rather, it is  argued that collapsed objects in cosmological spacetimes differ from the idealised Kerr or Schwarzschild black hole solutions with vacuum exteriors. Consequently, binary orbits can decay for two reasons -- firstly, the emission of gravitational radiation and secondly, as a consequence of the conservation of local angular momentum as the black hole masses increase.

\subsection{Orbital decay due to gravitational radiation}
Orbital decay via gravitational wave emission happens almost entirely in the weak field regime, which is tested empirically by binary pulsar observations \citep{Weisberg:2010zz}.   
We note that the cosmologically coupled black holes are claimed to be GEODEs, horizonless objects \citep{Croker_Weiner:2019} akin to   gravastars \citep{Berezin:1987bc,Dymnikova:1992ux, Mazur_Mottola:2001,Visser:2003ge,Mazur:2015kia}. However,  in this regime, the gravitational wave emission is  independent of the structure of near-point sources  \citep{Will:2011nz} and is governed by the time variation of the quadrupole moment tensor \citep{Peters_1964},
\begin{equation}
    Q_{i j}  = \Sigma_{a} m_{a} x_i x_j \,,
\end{equation}
where $\Sigma_{a}$ is the sum over the masses in the system located at positions $x_i$. The rate of energy loss due to  gravitational radiation  is 
\begin{equation}
     \left. \frac{dE}{dt} \right|_{GW}  =  \frac{G}{5c^5} \left(\frac{d^3 Q_{ij}}{dt^3} \frac{d^3 Q_{ij}}{dt^3} - \frac{1}{3} \frac{d^3 Q_{ii}}{dt^3} \frac{d^3 Q_{jj}}{dt^3}  \right) \,.
    \label{eq: power radiated}
\end{equation}

In our detailed calculations, we will assume the eccentricity ($e$) is zero; this underestimates gravitational wave emission, so our merger rates are effectively lower bounds. 
For a circular orbit, the energy emission rate is 
\begin{equation}
     \left. \frac{dE}{dt} \right|_{GW}   = -\frac{32}{5} \frac{G^4 \mu^2 M^3}{c^5 r^5} + {\cal{O}}(\dot m  )\,,
     \label{eq: energy loss due to GWs}
\end{equation}
where $M = m_1 + m_2$ and $\mu = m_1 m_2 / (m_1 + m_2)$ are the total and reduced masses of the binary, respectively.  Changing black hole masses  add  time derivatives to the above expression that are proportional to $\dot m$, $\ddot m$ and $\dddot m$. 
Differentiating Eq.~\ref{eq: Eqn from Farrah et al. for mass growth} gives
\begin{equation}
    \frac{dm}{dt} = kHm, 
    \label{eq: rate of growth of mass}
\end{equation}
where $H = \dot{a}/a$ is the Hubble parameter. Recalling that $\dot H \sim H^2$,  differentiating Eq.~\ref{eq: rate of growth of mass} gives $\ddot{m} \propto H^2 m$ and $\dddot{m} \propto H^3 m$. These contributions are thus far smaller than the time-independent $m$ term and are therefore ignored.
The total energy of the binary is
\begin{equation}
    E = -\frac{G\mu M}{2r} \,,
    \label{eq: potential energy}
\end{equation}
so gravitational wave emission causes $r$ to vary via the usual relationship
\begin{equation}
    {\left. \frac{d r}{d t}\right|}_{GW}  = \frac{2r^2}{G\mu M}    \left. \frac{dE}{dt} \right|_{GW}      = -\frac{64}{5} \frac{G^3 \mu M^2}{c^5 r^3} \, 
    \label{eq: orbital decay dr/dt}
\end{equation}
but with the masses now time-dependent. One can observe that as the Universe doubles in size, the mass of each black hole grows by $2^3$, boosting the overall  decay rate by $2^9=512$. Clearly, this will have a very substantial impact on merger dynamics.

\subsection{Orbital decay with fixed angular momentum }
The orbital angular momentum of an isolated binary with eccentricity $e$ is  
\begin{equation}
    L = \sqrt{G\mu^2 M r (1-e^2)} \,,
    \label{eq: AM of the binary}
\end{equation}
where $r$ now represents the semi-major axis. \cite{Croker:2020} argue that the eccentricity of a binary is not affected by mass growth. 

If $L$ is conserved outside of the loss due  to gravitational radiation, then $r$ will evolve as
\begin{equation}
    r = r_i \left(\frac{a}{a_i}\right)^{-3k}\, ,
    \label{eq: r evolution due to AM consevation}
\end{equation}
given that $\mu$ and $M$ grow with the masses in accordance with Eq. \ref{eq: Eqn from Farrah et al. for mass growth}.\footnote{As noted above, we set $e$ to zero in our detailed calculations. However, $M$  changes by orders of magnitude on cosmological timescales, so unless $e$ is initially very close to unity driving it to zero would only make a small change in $L$ in any case.}  Cosmologically coupled black holes gain mass by interacting with the overall spacetime, so it is not immediately obvious that \textit{local} angular momentum is rigorously conserved.  However, if we assume this to be true, then (where the subscript $AM$ indicates that the decay is calculated using Eq. \ref{eq: r evolution due to AM consevation})
\begin{equation}
    \left. \frac{dr}{dt} \right|_{AM} = - 3kHr \, .
    \label{eq: decay due to AM conservation}
\end{equation}
This would induce an additional correction to the time derivatives of the quadrupole moment tensor, but these are again suppressed by factors of $H$ and are thus ignorable. 

For a fixed angular momentum $L$, the rate of loss of energy can be calculated using Eq.~\ref{eq: potential energy}~and~\ref{eq: decay due to AM conservation} as
\begin{equation}
    \begin{aligned}
        & \left. \frac{dE}{dt}  \right|_{AM} = \frac{-G}{2} \frac{d}{dt} \left( \frac{\mu M}{r} \right) \\
        & = \frac{-G \mu M}{2r}  \left( 2kH - \frac{1}{r} \left. \frac{dr}{dt} \right|_{AM} \right) = \frac{-G \mu M}{2r} 5kH \,,
    \end{aligned}
    \label{eq: energy loss due to AM conservation}
\end{equation}
implying that in such a scenario the potential energy of a cosmologically coupled BBH is not conserved.

\subsection{BBH merger times with cosmological coupling} 


If we assume that the two energy loss mechanisms (Eq.~\ref{eq: energy loss due to GWs} and \ref{eq: energy loss due to AM conservation}) operate independently, then on differentiating Eq.~\ref{eq: potential energy}, the total rate of change in the separation of a circular BBH pair yields
\begin{equation}
   \frac{d r}{d t} = {\left. \frac{d r}{d t}\right|}_{GW} +  \left. \frac{dr}{dt} \right|_{AM} \, .
   \label{eq:loss_rate_combo}
\end{equation}
Defining
\begin{equation}
    \beta_i = \frac{64}{5} \frac{G^3 \mu_i M_i^2} {c^5} \,,
\end{equation}
where, as usual, the subscript $i$ indicates the ``birth'' value then
\begin{equation}
    \frac{d r}{d t}= - \frac{\beta_i}{r^3}\left(\frac{a}{a_i}\right)^{3 k}-3 k H r \, .
    \label{eq: decay rate due to both factors}
\end{equation}
The above can be written as
\begin{equation}
     \frac{d r^4}{d t}+12 k H r^4=-4 \beta_i\left(\frac{a} {a_i}\right)^{3 k} \,.
     \label{eq: ODE}
\end{equation}
Substituting $y(t) = r^4(t)$ yields a linear, first order  ordinary differential equation  
\begin{equation}
   \dot{y}(t) +P(t) y(t) =Q(t) \, ,
\end{equation}
where  $P(t) = 12kH$ and  $Q(t)=-4\beta_i ({a}/{a_i})^{3k}$. 
This  has a general solution (e.g., \citealt{Zwillinger:1989})
\begin{equation}
 y(t)= e^{- \int_{t_i}^{t} P(t')d t'} \left[\int_{t_i}^{t} e^{\int_{t_i}^{t'} P(t'') d t''} Q(t') dt'  + y(t_i)\right] \,, 
\end{equation}
where $t_i$ is the formation time of the BBH system.\footnote{We assume  the black holes  formed simultaneously, which is reasonable given the short lifetime and  presumed common origin of the progenitor stars.}  In addition, 
\begin{equation}
    e^{\pm \int_{t_i}^{t} P(t') d t'} = e^{\pm 12k \int_{t_i}^{t} \frac{\dot{a}}{a} dt'} = \left( \frac{a}{a_i} \right)^{\pm 12k} \,,
\end{equation}
so 
\begin{equation}
    y(t)=  \left(\frac{a}{a_i}\right)^{-12k} \left[y\left(t_i\right) -4 \beta_i \int_{t_i}^{t} \left(\frac{a}{a_i}\right)^{15k} d t' \right] \,.
\end{equation}
But $y(t) \rightarrow 0$ at merger, hence 
\begin{equation}
    \int_{t_i}^{T + t_i} \left(\frac{a}{a_i}\right)^{15 k} d t=\frac{y\left(t_i\right)}{4 \beta_i} = T_{k = 0} \,,
     \label{eq: solve for T}
\end{equation}
where $T_{k = 0}$ is the lifetime for  uncoupled black holes \citep{Peters_1964} and we have replaced the upper integration limit with $T+t_i$, $T$ being the lifetime of the cosmologically coupled BBH.

However, looking at Eq.~\ref{eq: decay rate due to both factors}, the initial decrease in $r$ is primarily driven by the requirement that $L$ is fixed while gravitational wave emission  dominates in the later stages. In fact, it is not fully clear whether $L$ is expected to be fixed in the presence of cosmological coupling – if the black holes are interacting with the overall spacetime, then in principle {\it local\/} angular momentum need not be fixed. Proponents of cosmological coupling would argue that this issue could be resolved by a suitable calculation. However, at this point, the conservation of $L$ for  BBH with time-varying masses is a hypothesis. Consequently, dropping the assumption of fixed $L$  such that the change in the separation is only due  to gravitational wave emission provides a  ``floor'' for estimates of the binary's lifetime. We distinguish these scenarios with $k$ and $k'$ - a primed value applies when the change in the separation is only due  to gravitational radiation.

\begin{figure}
    \vspace{-0.3cm}
    \centering
    \includegraphics[width = 1\linewidth]{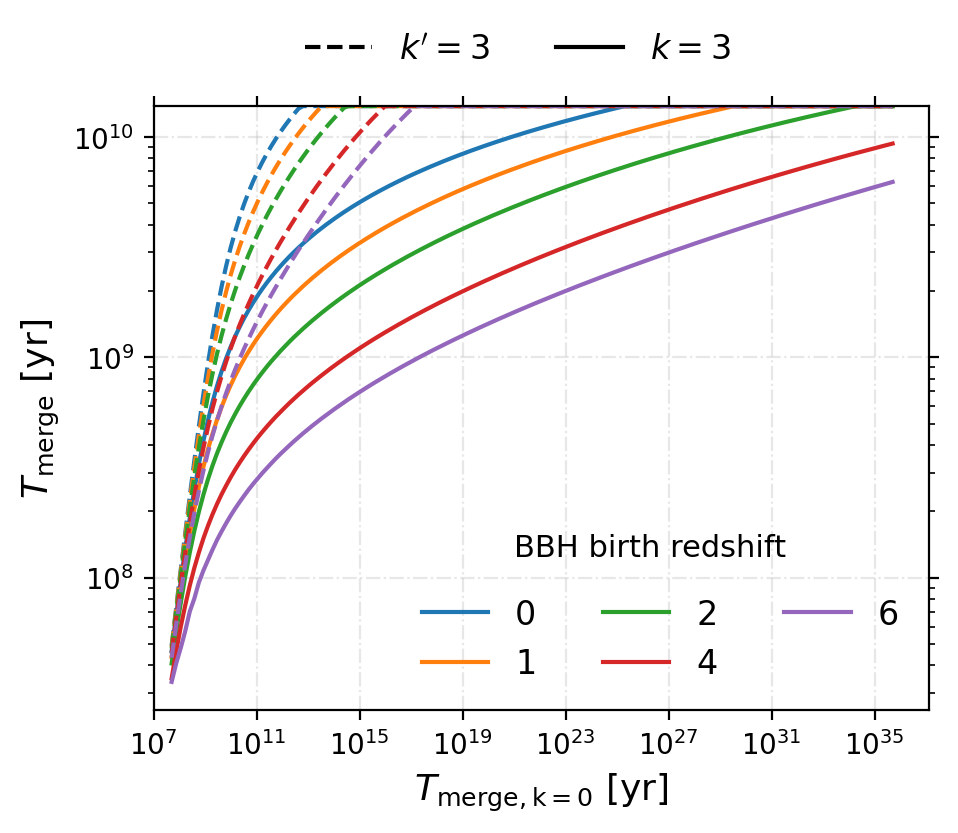}
    \caption{Merger times for cosmologically coupled black holes relative to the non-coupled ($k=0$) lifetimes. The line color represents the redshift at which the black holes  were formed. The solid lines show the $k = 3$ result with fixed angular momentum, and the dashed lines show the timescale with orbital decay mediated only by gravitational radiation, $k'=3$.}
    \label{fig: mergertime}
      \vspace{0.2cm}
\end{figure}

The merger time $T$ of the cosmologically coupled black holes can be found by integrating Eq.~\ref{eq: solve for T}. For a Friedmann-Robertson-Walker background, the cosmic time can be taken as 
\begin{equation}
    t(a) = \int_{0}^a \frac{da}{a H}  = \int_{0}^a \frac{da}{H_0 [\Omega_M a^{-1} + \Omega_{\Lambda} a^3]^{1/2}} \, ,
    \label{eq: cosmic time}
\end{equation}
where, as usual, $\Omega_M$ and $\Omega_\Lambda$ represent the present-day fractions of non-relativistic matter and dark energy, and $H_0$ is the current value of Hubble constant.\footnote{This is an approximation if  the dark energy density is sourced by cosmologically coupled stellar black hole population but it is sufficiently accurate for our purposes. We set $\Omega_\Lambda=  0.6925, \Omega_M = (1 - \Omega_\Lambda)$ throughout.} This gives
\begin{equation}
    a(t) = \left(\frac{\Omega_M}{\Omega_\Lambda}\right)^{1/3} \left[  \sinh \left(\frac{t}{\Tilde{t}} \right) \right]^{2/3}; \quad  \Tilde{t} = \frac{2}{3H_0 \Omega_\Lambda^{1/2}}\, .
    \label{eq: scale factor}
\end{equation}
We then evaluate the LHS of Eq.~\ref{eq: solve for T} to find
\begin{equation}
    \int_{t_i}^{T+t_i} \left(\frac{a}{a_i}\right)^{15k}  dt = \left(\frac{\Omega_M}{a_i^{3}\Omega_\Lambda}\right)^{5k} \int_{t_i}^{T+ t_i} \sinh \left(\frac{t}{\Tilde{t}} \right )^{10k} dt \,.
    \label{EQ: INTEGRATE ME}
\end{equation}
This integral can be performed analytically. However,  for $k=3$ the result is too tedious to quote explicitly.  Fig.~\ref{fig: mergertime} shows the resulting BBH merger times as a function of formation redshift, obtained by solving Eq.~\ref{eq: solve for T}.  

If we  consider orbital decay only gravitational wave emission, then the second term on the RHS  of Eq.~\ref{eq:loss_rate_combo} is not present and the equation can be integrated directly. In this case, the analogue of Eq.~\ref{eq: solve for T} is an integral over $(a/a_i)^{3k'}$. The time-dependence of $r$ is more complex when both decay mechanisms are active, but the lifetimes in the two scenarios are related by $k \leftrightarrow k'/5$. Hence, the $k'=3$ scenario in Fig. \ref{fig: mergertime} gives the same outcome as $k = 3/5$.

We can gain an intuitive sense of the impact of a coupling between black hole masses and cosmological expansion by assuming that the Universe is purely matter-dominated with $a/a_i = (t/t_i)^{2/3}$. In this case 
\begin{equation} 
T = t_i \left[ \left( \frac{(10k+1) T_{k = 0}}{t_i} + 1 \right)^{1/(10k + 1)} - 1 \right] \,.
\label{eq: Matter-dominated merger time}
\end{equation}
For example, a binary merging today, which formed at half the current age of the Universe, has its lifetime reduced by a factor of roughly $6.9 \times 10^7$ while the merger mass is eight times the birth mass. 
 
\begin{figure}
    \centering
    \includegraphics[width = 1\linewidth]{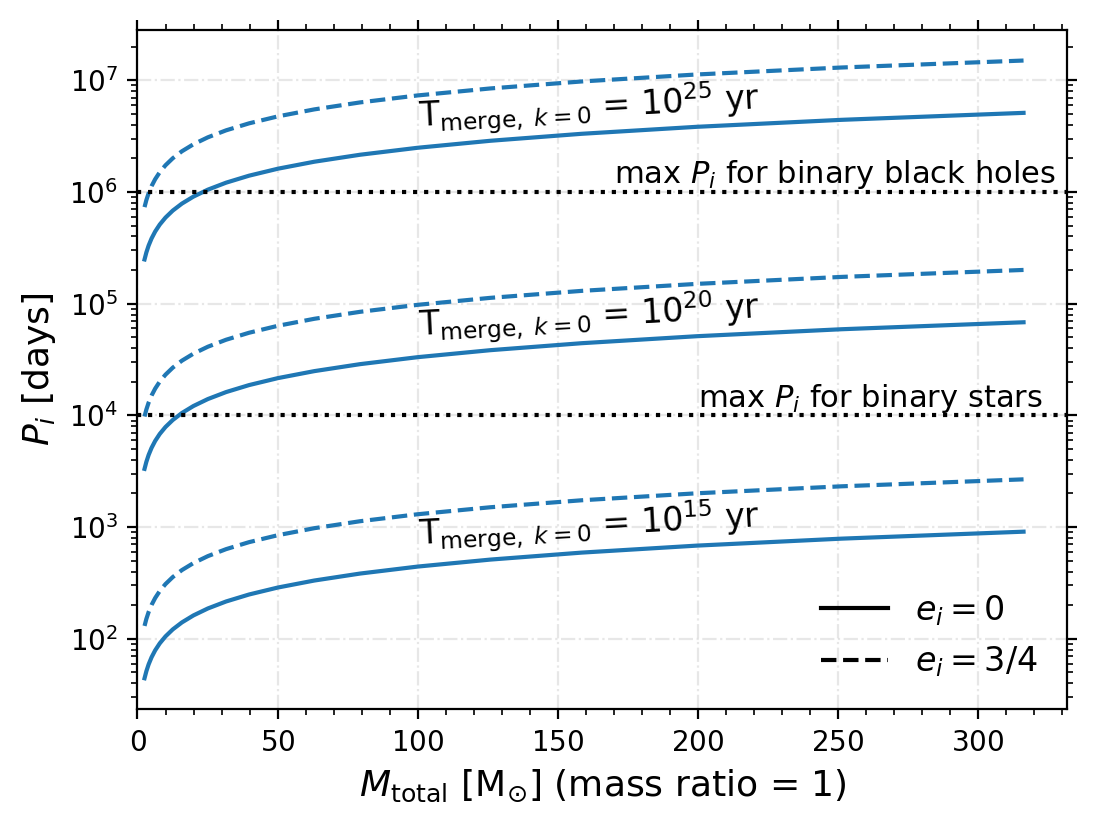}
    \caption{Merger time of equal-mass binary black holes as a function of their initial period $P_i$ and total mass $M_{\rm total}$ with $k = 0$. We give values for both initially circular and significantly eccentric ($e_i = 0.75$) systems. The dotted lines show the maximum values for the initial period of binary stars and black holes in our simulation.}
    \label{fig: mtot_vs_period}
    \vspace{0.3cm}
\end{figure}

Fig.~\ref{fig: mtot_vs_period} shows the expected lifetimes of binary systems in the standard scenario. The Universe is a little more than $10^{10}$ years old, so almost all binary systems would remain unmerged today. Cosmological coupling causes a dramatic decrease in merger time, boosting the merger rate and the merger masses will grow relative to standard expectations. 

Merger rates for BBH have been discussed previously by \cite{Croker_2021} and \cite{Farrah_2023}.  The former paper looks at the corresponding change in the merger rates of BBHs due to cosmological coupling with $k \leq 1$ and writes down an expression similar to Eq.~\ref{eq:loss_rate_combo} for the general elliptical ($e \neq 0$) case but the explicit calculation of the lifetime of circular binaries here is new.

\vspace{0.2cm}
\section{Expected binary black hole merger rate} \label{sec: merger_rates result}

\begin{figure}
    \centering
    \includegraphics[width = 1\linewidth]{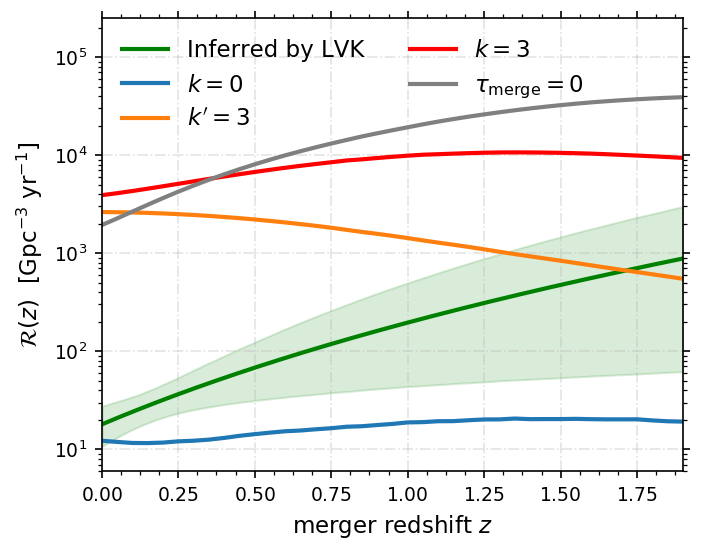}
    \caption{The source frame volumetric merger rate as a function of redshift for the three couplings strengths $k$ considered here. The green curve represents the  BBH merger rate as inferred by the LVK network, while the shaded area shows its 90\% credible interval (data taken from \citealt{GWTC_3_population_stats, GWTC_3_data_set}). The grey curve represents the rate under the assumption that the BBHs merge promptly on formation, i.e., their gravitational radiation-induced merger delay time $\tau_{\rm merge} = 0$.}
    \label{fig: R_volumetric_vs_z}
     \vspace{0.2cm}
\end{figure}

\begin{figure}
    \centering
    \includegraphics[width = 1\linewidth]{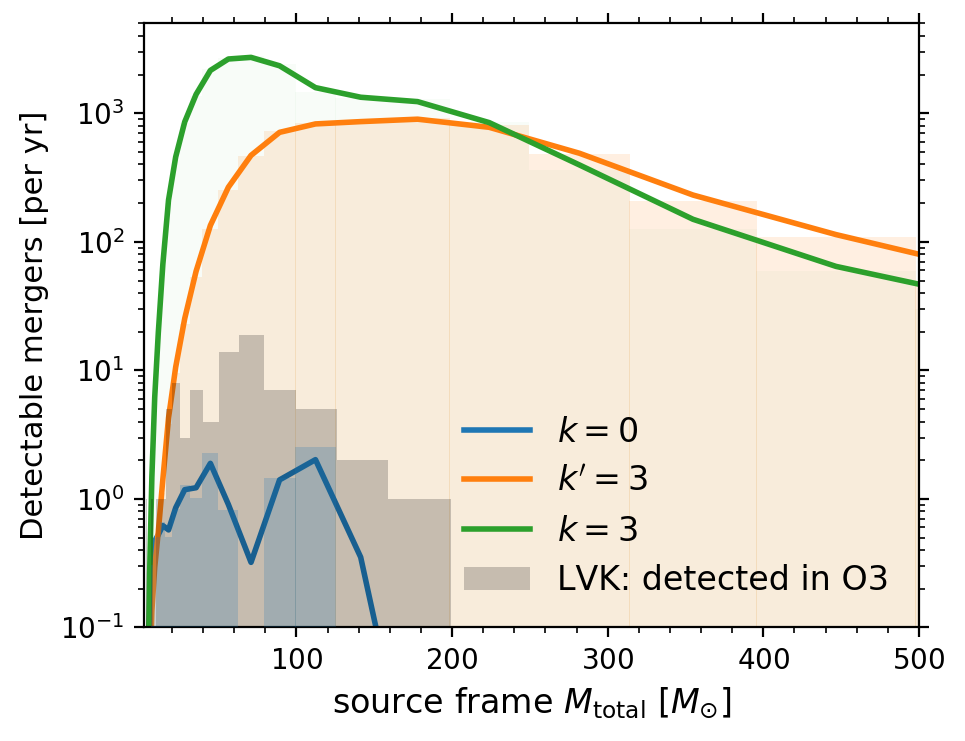}
    \caption{The expected number of detectable binary black hole mergers on earth (per mass bin, log spaced), given all three LVK detectors are working in quadrature for a full year at the third observing run (O3) sensitivity.}
    \vspace{0.2cm}
    \label{fig: detections per year}
\end{figure}

\begin{figure*} 
    \vspace{-0.2cm}
    \centering
    \includegraphics[width = 0.9\linewidth]{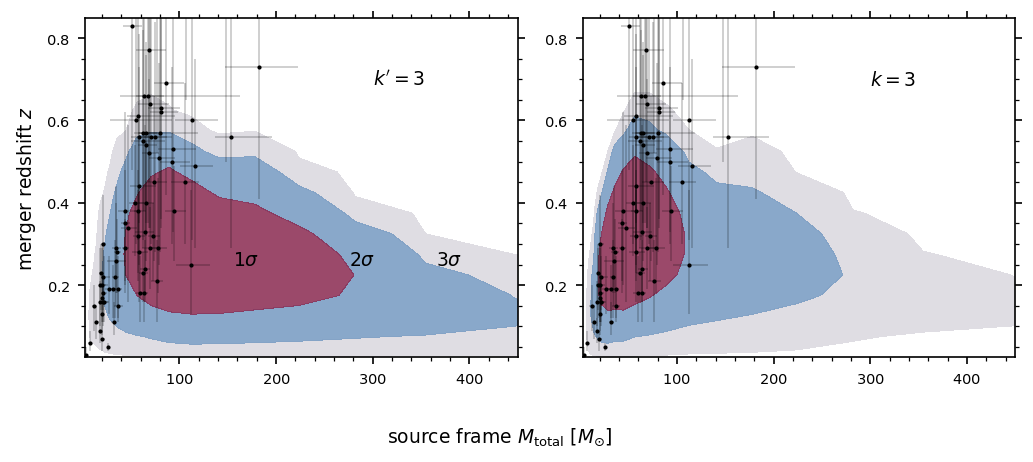}
    \includegraphics[width = 0.475\linewidth]{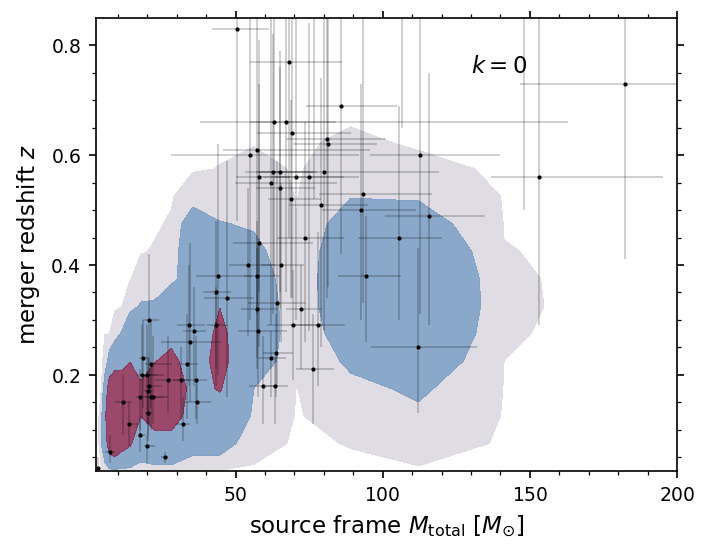}
    \caption{The parameter space distribution at $k^\prime = 3$ and $k = 3$ of the expected mergers on earth in a year under O3 sensitivity with that all the LVK detectors running for a full year. $M_{\rm total}$ represents the total binary mass at the time of the merger. The red, blue, and grey areas represent the region bounded by $1\sigma, 2\sigma$, and $3\sigma$ contour levels (contour level representing the expected rate). The overplotted data points are the observed values. The corresponding $k=0$ result is also included for reference and we allow  $e \neq 0$ in this case. }
    \label{fig: Mtot_vs_z_dect}
\end{figure*}

\cite{Farrah_2023} give a qualitative discussion of merger rates identifying the same shift toward shorter periods and higher masses that were found in the previous section. However, we now use  the explicit lifetime calculation obtained for circular BBH systems and apply it to a realistic initial population of BBH to compute the expected merger rate in the LVK detectors. 

We use \textsc{Bpass} v2.2.1 detailed stellar models (\citealt{Bpass2017, Bpass_2018}, also see \citealt{Bpass2019, Tang_2020, Briel_2022_a}) to generate the population of BBHs that are then distributed over a range of formation redshifts. These BBHs then undergo gravitational wave emission-induced merger. The details of population synthesis and merger rate calculation are discussed in Appendix~\ref{sec: merger_rate calc}.

The source frame  BBH merger rate density,\footnote{The source frame merger rate density at some redshift $z$ is the number of mergers happening in a Gpc$^3$ volume in a year (in source frame of reference) due to all the previously occurred star formation (see Appendix~\ref{sec: source frame rate}).}  along with the rate inferred from observations, is shown in Fig.~\ref{fig: R_volumetric_vs_z}. The $k=0$ simulation underpredicts the observed rate, which is attributable in part to \textsc{Bpass}  considering only isolated stellar binaries (see Appendix \ref{sec: popsyn}). However, with $k$ or $k'=3$, the rate is massively overpredicted, especially when $z \lesssim 1$.\footnote{Moreover, the $k=0$ \textsc{Bpass} prediction includes a range of initial eccentricities, unlike the $k>0$ results. An eccentric orbit would be more efficient in removing energy via gravitational radiation.}
Furthermore, when $k=3$, the BBHs merge rapidly, with most systems merging within a Hubble time. As a result, the merger rate shown in Fig. \ref{fig: R_volumetric_vs_z} closely aligns with the formation rate (grey curve), in contrast to the $k=0$ case. Consequently, the $k=3$ rate should be very weakly related to the $k=0$ one. However, if the difference in the implemented physics, such as common envelope evolution or mass transfer stability, affects the number of BBHs that can survive in the \textsc{Bpass} code compared to other population synthesis models,  and these BBHs happen to form with large enough periods such that they do not feature in the $k=0$ rate, then a discrepancy could arise between the $k'$ and $k=3$ rate predictions among studies employing different population synthesis codes.

Fig.~\ref{fig: R_volumetric_vs_z} also shows that the frequency of mergers peaks at $z\sim 1$ for $k=3$ but is still rising for $k'=3$. \cite{Farrah_2023} speculate that the accelerated merger rate means that cosmologically coupled BBH combine at redshifts high enough to make them undetectable, but we find that the merger rate today is still nontrivial. 

To calculate the expected detection rate, we assume that all three LVK detectors are operating in quadrature at the third observing run sensitivity with threshold signal-to-noise ratio for detection $\rho = 8$ (see Appendix \ref{sec: detectable rate}).
The predicted number of detectable mergers for the LVK network is shown in Fig.~\ref{fig: detections per year}. As expected, there is a great discrepancy between this and the observed merger count with both $k'$ and $k = 3$. 

The total mass versus merger redshift distribution is shown in Fig.~\ref{fig: Mtot_vs_z_dect}. Most of the  ``massive" BBH mergers occur at lower redshifts -- these black holes would have been born far apart at higher $z$ and their mass increases significantly before the merger takes place.  Enforcing angular momentum conservation ($k=3$) leads to more rapid orbital decay and, therefore, more frequent mergers than gravitational wave emission alone ($k^\prime = 3$). Since black holes merge more rapidly in the former scenario, the merger mass is reduced relative to the $k^\prime = 3$ case.  

Seismic noise decreases the sensitivity of the LVK detectors below 10 Hz. Since more massive binaries generate lower frequency signals, these systems are more visible at lower redshifts because both the amplitude and frequency of the detected signal will be larger. This is the reason why the contours in Fig. \ref{fig: Mtot_vs_z_dect} become concentrated towards lower redshifts at larger binary masses.

Binary population studies contain many layers of uncertainties that could affect the BBH merger rate (see, e.g., \citealt{Broekgaarden_2022}). However, the raw result \textsc{Bpass} (i.e., $k = 0$) is a tolerable fit to the detected mass distribution. Conversely, the dramatic increase in orbital decay for coupled black holes (see Figs.~\ref{fig: mergertime} and \ref{fig: mtot_vs_period}) means that their predicted merger rate exceeds observations by several orders of magnitude. In addition, due to the mass growth between formation and merger, the LVK network should have detected a hundred or more events with combined masses in excess of 200 M$_\odot$ but has seen none. Consequently, we find a dramatic disagreement between observational data and predictions made if black hole masses vary fast enough to explain the apparent SMBH growth in red sequence ellipticals \citep{Farrah_2023a}.

\section{Spin and cosmological coupling} \label{sec: spin of cosmological coupled BH}

The dimensionless Kerr spin parameter, $a_{\rm BH}$, of a black hole with mass $m$ is
\begin{equation}
    a_{\rm BH} = \frac{Jc}{G m^2} \, ,
    \label{eq: dimensionless spin}
\end{equation}
where $J$ is the spin angular momentum of the black hole. If cosmologically coupled black holes gain mass without also gaining angular momentum, then 
\begin{equation}
     a_{\rm BH}(t) =  a_{{\rm BH},i} \left(\frac{ a(t)}{a_i}\right)^{-2k} \, ,
     \label{eq: BH spin}
\end{equation}
where $a_{ {\rm BH},i }$ is the value of the spin parameter at $t_i$. 

Table~\ref{tab: a_* values} shows the spin parameters of three SMBHs (residing in elliptical galaxies) used by \cite{Farrah_2023a} for which there are measured values and the affirmatively measured values are close to unity.  By hypothesis SMBHs -- or at least those in red sequence ellipticals -- are not being spun up by accretion since this would undermine the argument for cosmological coupling. Likewise, major mergers are rare at $z \lesssim 1$.\footnote{In principle, the SMBH merger may occur well after the merger of their original host galaxies. These mergers are invisible to the LVK network but will be explored by LISA \citep{LISACosmologyWorkingGroup:2022jok}.}

\begin{table}
    \centering
    \caption{Kerr parameter values for \cite{Farrah_2023a} SMBHs}
    \label{tab: a_* values}
    
    \begin{tabular}{ccc} 
    \hline
    
     Galaxy  &  $a_{\rm BH}$  & Source \\
    \hline \hline
        M87 &  $0.90 \pm 0.05$   & \cite{Tamburini_2020} \\
        Mrk 509 &   $0.84 \pm 0.01$ & \cite{Piotrovich_2017} \\
        3C 273 &  $< 1$& \cite{Piotrovich_2017} \\
    \hline
    \end{tabular}
    \vspace{0.5cm}
\end{table}

Horizonless compact objects can have $a_{\rm BH} > 1$ without a naked singularity.  However, the spin parameter of many black holes has been inferred from observations but none have been found with values exceeding unity. Consequently, without significant accretion at $z\lesssim 1$ we expect that an SMBH that initially had  $a_{{\rm BH}, i} \approx 1$  would have $a_{\rm BH} \approx 0.016$ (for $k=3$) at $z = 0$, via Eq.~\ref{eq: BH spin}.  It is possible that cosmological coupling leaves $a_{\rm BH}$ fixed rather than $J$ but this line of reasoning does have the potential to constrain detailed proposals.

\section{Minimum mass of primordial black holes } \label{sec: minimum mass of primordial BHs}

The collapse of primordial overdensities \citep{1967SvA....10..602Z,Hawking_1971} or small-scale nonlinear structures \citep{1989PhLB..231..237H,Green:2000he,Eggemeier:2021smj} can lead to the formation of PBHs.  Small black holes lose mass via Hawking radiation and if their initial mass is less than $\sim 10^{15} $~g they would evaporate completely in the current lifetime of the Universe.  

Hawking radiation is less dependent on the actual presence of an horizon than is often realized \citep{Visser:2001kq,Bardeen:2014uaa}.  Consequently, we can use the usual expressions for these quantities in this initial exploration and a cosmological coupling will stabilize a PBH if its mass gain rate, described by Eq.~\ref{eq: rate of growth of mass}, exceeds the loss rate due to Hawking radiation \citep{Hawking_1975},
\begin{equation}
    \frac{dm}{dt} = -\frac{\hbar c^4}{15360 \pi G^2} \frac{1}{m^2} \, .
    \label{eq:Hawking Mdot}
\end{equation}

During radiation domination 
\begin{equation}
    H^2 = \frac{8 \pi G}{3}  \frac{g\pi^2}{30} \frac{(e T)^4}{(\hbar c)^3} \, ,
\end{equation}
where $g$ is the number of effective degrees of freedom, $e$ is the charge on the electron, and $T$ is the temperature of the Universe in eV. 
Equating Eq.~\ref{eq: rate of growth of mass} and~\ref{eq:Hawking Mdot} gives the minimal mass of a stabilized black hole in terms of $T$, 
\begin{equation}
    m_i \ge \frac{1}{8} \left( \frac{c^{11} \hbar^5}{80 e^4 g G^5 k^2 \pi^5 T^4} \right)^{1/6} \approx \frac{2.37\times 10^6}{T^{2/3}}   \, \mathrm{kg} \,.
    \label{eq:balance mass}
\end{equation}
We fix $g=2$ (thus counting only photon modes) for convenience, noting that the dependence on the precise value is very weak. The present day mass of a black hole that saturates the  above condition is   
\begin{equation}
    m \approx 2.37\times 10^6 \frac{T^{7/3}}{T_{\mathrm{CMB}}^3}  \, { \rm kg} \, 
\end{equation}
with $k=3$, and we quantified the growth of the Universe since the formation of the PBH via $a\sim 1/T$, again assuming that $g$ is fixed. 

Given $T_{\mathrm{CMB}} = 6.6 \times 10^{-4}$~eV, a black hole produced during nucleosynthesis   ($\sim 0.1$ MeV)  with  $m_i \sim 10^3$~kg would be stable, with a present-day mass of around $0.001 M_\odot$.  However, black hole production during nucleosynthesis is likely to be disruptive to nucleosynthesis itself. Less constrained formation mechanisms at  TeV scales and above lead to  PBH with a present day masses of at least $4\times 10^{11} M_\odot$, which are infeasibly large.  Consequently, the existence of a primordial population of black holes is unlikely to be consistent with a cosmological coupling like that proposed by \cite{Farrah_2023} unless they form in the relatively late universe. For example,  \cite{Chakraborty:2022} describes a scenario where PBHs with mass $10^{-16} M_\odot$ form at $z\sim 10^6$. These objects would have present day masses of $100 M_\odot$ but would substantially overclose the Universe for almost any nontrivial initial mass fraction.

\section{Discussion} \label{sec: Discussion}

We consider the astrophysical implications of cosmologically coupled black holes whose masses grow with the third power of the scale factor, keeping their density constant, as envisaged by \cite{Farrah_2023}. We consider the expected merger rates of stellar black holes, the evolution of the black hole spin parameters in galaxies devoid of active accretion, and the minimum mass of current epoch PBH.

Our key result is that a cosmological coupling that mimics the dark energy component of the Universe would yield a BBH merger rate inconsistent with that seen by the LVK network.  Specifically,  the number of detections rises by several orders of magnitude relative to the uncoupled case, with their typical masses increasing by a factor of several and many events expected above $200M_\odot$. 

We work with the \textsc{Bpass} population synthesis tool in conjunction with the star formation rate density of Eq.~\ref{eq: SFRD_langer_norman}. The  \textsc{Bpass} population underpredicts the merger rate relative to LVK  but is known to neglect some formation channels, as discussed in Section~\ref{sec: popsyn}.  In addition, we ignore the impact of eccentricity on BBH lifetimes when $k \neq 0$, further reducing the computed merger rate. Consequently, our predicted merger rates for coupled black holes are a conservative lower bound. Accommodating the merger rate induced by cosmological coupling would thus require a massive revision of binary stellar evolution which is separately constrained by multiple astrophysical considerations. 

We did not consider a range of $k$-values but our results are qualitatively consistent (though quantitatively different) with an extrapolation of the findings for a weaker coupling strength of $k \leq 1$, analyzed by \cite{Croker_2021}. They likewise see that a cosmological coupling produces massive black holes that are otherwise absent in traditional stellar evolution models. 

Separately, in Section \ref{sec: spin of cosmological coupled BH}, we show that the Kerr spin parameter of the SMBH in matter-deficient elliptical galaxies would naively satisfy the condition set by Eq. \ref{eq: BH spin} and $ a_{\rm BH}(t) =  ({ a(t)} / {a_i})^{-2k}$. However, the handful of SMBH  in the dataset of \cite{Farrah_2023a} for which the spin has been inferred are spinning much faster than expected if $k = 3$ and $a_i \lesssim 1$. 

In Section~\ref{sec: minimum mass of primordial BHs}, we show that any surviving PBH should grow to a mass $\gtrsim 0.001 M_{\odot}$ today and that those forming at genuinely early times will have masses similar to large galaxies. Obviously, PBHs are so far hypothetical, but it is interesting to see that any such population would be dramatically changed by a cosmological coupling.

The underlying motivation for the hypothesis considered here was derived from considerations involving black holes embedded directly into  expanding spacetimes. However, almost all known astrophysical black holes (whether of stellar origin or SMBH) are located inside much larger galactic potential wells and thus interact minimally with the  Hubble flow. The impact of this situation on the underlying theoretical motivation is unclear \citep{Wang:2023}. Thus it may not immediately follow that increasing the masses of objects inside non-expanding regions of spacetime would mimic a cosmological constant with sufficient fidelity to be consistent with current constraints on the properties of dark energy. Moreover, the masses of isolated galaxies would increase with time in this scenario,  given that most black holes are found inside galaxies. This will presumably have significant, unexamined consequences for the internal dynamics of galaxies and alter expectations for the BBH population in the Milky Way \citep{2022ApJ...937..118W}.

Finally, this situation serves as an interesting case study of attempts to explain apparently anomalous observational results (in this case, the apparent growth of SMBH in galaxies unable to support accretion) by invoking changes to conventional understandings of fundamental physics. Such hypotheses are key to progress in fundamental science, but their wider astrophysical consequences must be considered when evaluating the robustness of such ideas.

\section*{Acknowledgements} 
\noindent 
We thank the anonymous referee for feedback on the manuscript. SG is supported by the University of Auckland doctoral scholarship. RJME, JJE and MMB acknowledge support of Marsden Fund Council managed through Royal Society Te Apārangi. This project utilized NeSI high-performance computing facilities.  We thank Kevin Croker and Duncan Farrah (and  collaborators) for valuable correspondence and a careful reading of an earlier draft of this work.

\bibliographystyle{mnras}
\bibliography{ref} 

\vspace{-1cm}
\appendix

\section{Merger rate calculation} \label{sec: merger_rate calc}

\subsection{Stellar population synthesis} \label{sec: popsyn}
 
Using the \textsc{Bpass} v2.2.1 detailed stellar models \citep{Bpass2017, Bpass_2018} we simulate a population of massive stars that are born in binaries with a range of initial masses ($m_{1, i} \leq 300$ M$_{\odot}$), mass-ratios ($q_i:= m_{2, i}/m_{1, i} \in [0.1 - 0.9]$),  periods ($P_i \leq 10^4$ days) and metallicities ($Z \in [10^{-5} - 0.04$]) respectively. The $q_i$ and $P_i$ distributions are based on the observational findings of \cite{Moe_Di_Stefano2017}, namely that the $q_i$ follow a flat distribution, i.e.,
\begin{equation}
    p(q_i) = 1 \,,
\end{equation}
$p$ being the probability distribution function and the $P_i$ obey a flat-in-log distribution (independent of $m_i$)
\begin{equation}
    p(P_i) \propto 1/ P_i \,.
    \label{eq: period distribution}
\end{equation}
This implies that there are fewer binaries at relatively large periods. The maximum initial period for stellar binaries is set to $P_i \leq 10^4$ days but we allow black hole binaries with periods of up to $10^6$ days, given that periods can be increased by mass-loss and supernovae kicks. The initial eccentricity is assumed to be zero.

\textsc{Bpass} assumes an instantaneous starburst of a fixed amount of matter with the formation of stars with various masses and periods.  Their abundance is determined by the initial mass function proposed by \cite{Kroupa2001},
\begin{equation}
    p(m_{1, i}) \propto m_{1,i}^{-2.3} \,.
\end{equation}
As the binary evolves, episodes of mass transfer can harden the system\footnote{Mass accretion in \textsc{Bpass} is only limited by the thermal timescale of the mass-gaining star. Unimpeded super-Eddington accretion is allowed if the mass gainer is a black hole (e.g., \citealt{Briel_2022_b}). Mass accretion episodes may also lead to so-called chemically homogeneous evolution (e.g., \citealt{Maeder2000}), and we implement such evolution in \textsc{Bpass} following \cite{Ghodla_2023}. We also consider tidal locking-induced homogeneous evolution in binary counterparts (e.g., \citealt{deMink2009, deMink_Mandel2016, Marchant2016}). But since \textsc{Bpass} does not have binaries with mass-ratio unity, we underestimate the number of homogeneous stars resulting from tidal locking.}. Eventually, nuclear fusion ceases to support the stars, and if they are massive enough\footnote{We set the maximum mass of a neutron star as 2.5 M$_{\odot}$.} we are left with a black hole binary. To calculate the subsequent gravitational wave emission-induced merger time, we follow \cite{Peters_1964} using the fits of \cite{Mandel_fit_2021}.

There is a possibility of a supernova (SN) kick which can alter the binary's period and eccentricity or even unbind the system. We account for SN kicks by using the Maxwellian distribution described in \cite{Hobbs_2005}. We also allow for the occurrence of $e^{-}e^{+}$ pair-instability SN \citep{Fowler_and_Hoyle1964} in stars with helium core mass $\in [65-130]$ M$_{\odot}$. In these cases, the system totally disrupts, leaving no remnant behind. Slightly less massive stars on the LHS of the pair-instability mass range may still undergo pair production but not in a sufficient amount to unbind the star. Consequently, these stars might experience a few oscillatory pulses leading to mass ejection episodes prior to their SNe explosion \citep{Woosley_PPISNe_2007}. Pulsation pair-instability SN is taken into account by following the prescription in \cite{Stevenson_2019} for non-rotating helium star models of \cite{Marchant_2019_PPISne}.

Gravitational wave mergers could also arise via the dynamical capture of black holes in dense astrophysical environments (e.g., \citealt{Kulkarni_1993}), three-body interactions (e.g., \citealt{Ziosi:2014}) or the efficient inspiral of compact remnants in the disks of active galactic nuclei (e.g., \citealt{McKernan_2012}). However, here we only consider BBH mergers originating from isolated binary stellar evolution which can lead to an undercount. 
There are a large number of uncertainties in the theory of single and binary stellar evolution. These include the treatment of convection (e.g., \citealt{Kupka_2017}), rotation (e.g, \citealt{Langer_2012_review_massive_stars}), stellar winds (e.g., \citealt{Vink_mdot_WR_2005, Brott2011}), mass transfer (e.g., \citealt{Ivonova2013}), SN explosion (e.g., \citealt{Muller_2016_3D}) followed by the natal kick (e.g., \citealt{bray2018neutron, Mandel_Muller, Ghodla2022, Richards_2022}), remnant mass calculation (e.g., \citealt{Fryer2012, Mandel_Muller, Ghodla2022}), among others. We treat these in the standard fashion detailed in \cite{Bpass2017}.

For our purposes, the relevant output of the \textsc{Bpass} simulation is recorded in the $\mathcal{R}_{\rm{smpl}}(m_1, m_2, \tau, Z)$ variable. It contains the merger rate density of the  BBHs, per unit mass of star formation in the simulation as a function of the component masses ($m_1, m_2)$, the total delay time $\tau$ (i.e., time from starburst to  BBH merger) and progenitor metallicity. 

\subsection{Source frame volumetric merger rate for $k=0$} \label{sec: source frame rate}

Let us assume that the total delay time from the onset of a binary star formation to its eventual gravitational radiation emission-induced merger is $\tau$. Then the source frame  BBH merger rate density $\mathcal{R} (m_1, m_2, t, Z)$ at some time $t$ due to all the previously occurred star formation at metallicity $Z$ is 
\begin{equation}
    \mathcal{R}_{k=0} (m_1, m_2, t, Z) =\int_{\Tilde{t}}^{t} \psi(t-\tau, Z) \cdot \mathcal{R}_{\rm{smpl }}(m_1, m_2, \tau, Z) d \tau \,.
    \label{eq: R_vol as a function of metallicity}
\end{equation}
Here, $\psi(t-\tau, Z)$ is the star formation rate density (SFRD) at metallicity $Z$ and $\mathcal{R}_{\rm smpl} (m_1, m_2, \tau, Z)$ is the merger rate density of BBHs resulting from $\psi$ with merger delay time $\tau$. Also, the lower limit of integration $\Tilde{t}$ marks the beginning of star formation and is taken as the time when the Universe was $\sim 200$ Myr old. The $\mathcal{R}_{\rm smpl}$ function is generated by the \textsc{Bpass} simulation (sec. \ref{sec: popsyn}).  The form of $\psi$ as a function of redshift is taken from \cite{Madau_Dickinson2014}.
 
To account for the chemical enrichment history of the Universe, we combine $\psi$ with the analytic expression for metallicity evolution from \cite{Langer_Norman_2006}. Consequently, at a given redshift $z$, the cumulative SFRD as a function of metallicity $Z$ is given by
\begin{equation}
    \Psi \left(z, Z / Z_{\odot} \right) = 0.015 \frac{(1+z)^{2.7}}{1+[(1+z) / 2.9]^{5.6}} \times \frac{\hat{\Gamma}\left[0.84,\left(Z / Z_{\odot}\right)^{2} 10^{0.3 z}\right]}{\Gamma(0.84)}  \;\;\; [\mathrm{M}_{\odot} \mathrm{Mpc}^{-3} \mathrm{yr}^{-1}],
    \label{eq: SFRD_langer_norman}
\end{equation} 
where $Z_{\odot}$ is the solar metallicity (=0.02) and $\hat{\Gamma}$ and $\Gamma$ are the incomplete and complete Gamma functions. From this definition of $\Psi$, we then reconstruct the form of $\psi$ defined in Eq. \ref{eq: R_vol as a function of metallicity}.
Finally, integrating Eq. \ref{eq: R_vol as a function of metallicity} over $Z$ gives the source frame merger rate density 
as  
 \begin{equation}
     \mathcal{R}_{k=0} (m_1, m_2, t) =\int_{0}^{\infty} \mathcal{R}_{k=0} (m_1, m_2, t, Z) \, d Z \; \; [{\rm Mpc}^{-3}{\rm yr}^{-1}] \,.
\end{equation}
Changing the SFRD and metallicity evolution in Eq. \ref{eq: R_vol as a function of metallicity} would directly influence the hence calculated merger rate density (e.g., \citealt{Tang_2020, Briel_2022_a}).

\subsection{Source frame volumetric merger rate for $k \neq 0$}

To illustrate the calculation analytically, we assume a purely matter-dominated Universe. One can then extend this analysis numerically to the more general case. With this assumption, the change in the merger times between the coupled and the non-coupled scenario can be calculated using Eq. \ref{eq: Matter-dominated merger time}. For consistency of notation, we rewrite this equation in terms of $\tau$,
\begin{equation} 
\tau_{k \neq 0} = t_i \left[ \left( \frac{(10k+1) \tau_{k = 0}}{t_i} + 1 \right)^{1/(10k + 1)} - 1 \right] \,.
\end{equation}
In a differential form this becomes
\begin{equation}
     d\tau_{k=0} =  { \left[ \frac{\tau_{k \neq 0}}{t_i} + 1 \right]^{10k}} d \tau_{k \neq 0} \,.
\end{equation}
For $k \neq 0$, $\mathcal{R}_{\rm{smpl }}(m_1, m_2, \tau, Z) d \tau $ is thus
\begin{equation}
    \mathcal{R}_{\rm{smpl}} \left(m_1, m_2 , \frac{t_i}{10k +1} \left[ \left( \frac{\tau_{k \neq 0}}{t_i} + 1 \right)^{10k + 1} - 1 \right], Z  \right) { \left[ \frac{\tau_{k \neq 0}}{t_i} + 1 \right]^{10k}} d \tau_{k \neq 0} \,.
\end{equation}
We note that although the function $\mathcal{R}_{\rm smpl}$ remains the same, its domain spanned by the integral in Eq. \ref{eq: R_vol as a function of metallicity} has changed and now allows for mergers with much larger delay time.
Additionally, the delay between star formation time and the formation time of the BBH needs to be calculated within the simulation. For the sake of demonstration, if here we have simply considered this to be zero, then  $t_i = t - \tau_{k \neq 0}$.
Therefore, for $k \neq 0$, Eq. \ref{eq: R_vol as a function of metallicity} is transformed as
\begin{equation}
    \mathcal{R}_{k \neq 0}(m_1, m_2, t, Z) =\int_{\Tilde{t}}^{t} \psi(t-\tau_{k \neq 0}, Z) \cdot \mathcal{R}_{\rm{smpl}} \left(m_1, m_2, \frac{t - \tau_{k \neq 0}}{10k +1} \left[ \left( \frac{t}{t - \tau_{k \neq 0}} \right)^{10k + 1} - 1 \right], Z  \right) \frac{d \tau_{k \neq 0}}{\left[ \frac{t}{t - \tau_{k \neq 0}} \right]^{-10k}} \,.
\end{equation}

\subsection{Observer frame intrinsic merger rate}  \label{sec: intrinsic rate}

The maximum possible rate observed on Earth (i.e., at infinite detector sensitivity) can be calculated as
\begin{equation}
    \mathcal{R}_{\rm{intr}} (m_1, m_2) =\int_{0}^{\Tilde{z}} \frac{\mathcal{R} (m_1, m_2, t_{z})}{1 + z} d V(z) \;\; [{\rm yr}^{-1}] \,.
\end{equation}
From here on, we drop the $k$ subscript from the rates as the calculation would be analogous for both $k =0$ and $k \neq 0$.
We set $\Tilde{z}$, corresponding to the time $\Tilde{t}$ in Eq. \ref{eq: R_vol as a function of metallicity}. The $1/(1+z)$ factor scales the source time to observer time (i.e. $t$ at $z = 0$) and $dV$ is a differential comoving spatial volume element:
\begin{equation}
    d V(z) =\frac{4 \pi c}{H_{0}} \frac{D_{c}^{2} (z)}{E(z)} d z \,.
\end{equation}
Also $E(z) =\sqrt{\Omega_{M}(1+z)^{3} + \Omega_{\Lambda}}$, $H_0$ is today's value of Hubble's constant and 
\begin{equation}
    D_{c}(z) =\frac{c}{H_{0}} \int_{0}^{z} \frac{d z^{\prime}}{E\left(z^{\prime}\right)} 
\end{equation}
is the comoving distance.
It is assumed that we are well into the matter-dominated epoch in a flat $\Lambda$CDM cosmology\footnote{A fair assumption as stellar black holes could not have formed earlier.} with $\Omega_\Lambda = 0.6925$ and $H_0 = 67.9$ km s$^{-1}$Mpc$^{-1}$. This choice of the cosmological parameters matches those in \cite{GWTC_3_population_stats} and is based on the work of the \cite{Planck_Collaboration_2016} and we map the SFRD to suit these values. We refer to $\mathcal{R}_{\rm intr}$ as the intrinsic merger rate.

\subsection{Detection rate}  \label{sec: detectable rate}

To determine the detectable rate from the above-described intrinsic rate, the detection probability of each merger event needs to be calculated. For this, we use the gravitational waveform synthesis tool \textsc{Riroriro} \citep{riroriro} and provide a brief summary of the methodology in the following. Assuming an isotropic source distribution (valid at the cosmological scale considered here), the probability of detection $p_{_d}$ for each merger event in $\mathcal{R_{\rm intr}}$ is 
%
\begin{equation}
    p_{_d}(z, m_1, m_2)= 1- C_{\rm{CDF}} \left[\min \left(\frac{8}{{\rho}_{\rm opt}(z, m_1, m_2)}, 1\right)\right]  \,,
\label{eq: detection_probability}
\end{equation} where $\rho_{\rm opt}$ is the signal-to-noise ratio of a gravitational waveform at the location of the detector, assuming optimum alignment and takes the form
\begin{equation}
    {\rho}_{\rm opt} = 2 \left[  \int_{f_{\min }}^{f_{\max }}  \frac{|\Tilde{h}(f)|^2}{S(f)} df \right]^{1/2}.
\end{equation} Here $f, \Tilde{h}$ represents the redshifted gravitational wave frequency and the redshifted and Fourier-transformed gravitational waveform of the BBH, assuming the source to be non-spinning, face-on, and perfectly aligned with the detectors. Considering spin would lead to a larger value of $\rho_{\rm opt}$ as such, we might underestimate the detection rate for certain systems. The more massive black holes would have a lower merger frequency. Thus, we set $f_{\rm min} = 1$ Hz while $f_{\rm max}$ cutoff is set near the conclusion of the ringdown phase.
$S(f)$ is the power spectral density of the noise in a detector, where we use the O3-run sensitivity noise curve.

We utilize a lower limit of $\rho = 8$ for detectability.
This is in contrast to the $p_{\rm astro} > 1/2$ criteria (probability of the gravitational wave signal being of astrophysical origin)  used by LVK for a secure detection which might, in some cases, lead to a confident detection even when $\rho < 8$. 

The projection function $\Theta$ is used to consider the effect of the arbitrary alignment of the source and the detector on detectability  \citep{Finn_1996}. A  distribution of $\Theta$ for a random source/detector alignment is calculated. Then its cumulative distribution function $C_{\rm CDF}$ is used in Eq. \ref{eq: detection_probability} to calculate the probability of orientations for which a merger is detectable (e.g., \citealt{Belczynski2014}). 
The subtracted term in the Eq. \ref{eq: detection_probability} gives the proportion of $\rho_{\rm opt}$ weighted with the projection function $C_{\rm CDF}$ which have $\rho < 8$ and hence considered undetectable. \\


\label{lastpage}
\end{document}